\documentclass[smallextended]{svjour3}       
\smartqed  
\usepackage{amssymb,graphicx}

\def\be{\begin{equation}}
\def\ee{\end{equation}}
\def\bea{\begin{eqnarray}}
\def\eea{\end{eqnarray}}
\def\grad{\mathbf{\nabla}}

\def\v{\mathbf{v}}
\def\x{\mathbf{x}}

\def\U{\mathbf{U}}

\def\J{{\cal J}}

\def\P{{\cal P}}
\def\E{{\cal E}}
\def\L{{\cal L}}
\def\T{{\cal T}}
\def\R{{\cal R}}

\def\pd{\partial}
\def\d{\mathrm{d}}

\def\at{\tilde{\alpha}}

\def\exp{\mathrm{exp}}

\def\etal{\textit{et.al.} \,}
\def\ie{\textit{i.e.} \, }
\def\rp{{r^\prime}}
\def\rs{{r_\mathrm{o}}}

\begin{document}

\title{Gluon matter plasma in the compact star core within fluid QCD model}

\author{C.S. Nugroho$^{a}$ \and A.O. Latief$^{a}$ \and T.P. Djun$^{b}$  \and L.T. Handoko$^{a,b}$}

\institute{$^{a}$ Department of Physics, University of Indonesia, Kampus UI Depok, Depok 16424, Indonesia
           \at
           $^{b}$ Group for Theoretical and Computational Physics, Research Center for Physics, Indonesian Institute of Sciences, Kompleks Puspiptek Serpong, Tangerang 15310, Indonesia
}

\date{Received: date / Accepted: date}

\maketitle

\begin{abstract}
The structure of compact star core filled by gluon matter plasma is
investigated within the fluid-like QCD framework. The energy momentum tensor,
density and pressure relevant for gluonic plasma having the nature of fluid
bulk of gluon sea are derived within the model. It is shown that the model
provides a new equation of state for perfect fluid with only a single parameter
of fluid distribution  $\phi(x)$.  
The results are applied to construct the equation of state describing the
gluonic plasma dominated compact star core. The equations of pressure and
density distribution are solved analytically for small compact star core radius.
The phase transition of plasma near the core surface is also discussed.
\end{abstract}

\section{Introduction}
\label{sec:intro}

Recent experiments in the last decades on relativistic nuclear collisions shed light on the phenomena of hot plasma formed by dense quarks and gluons. Those experiments suggest that the quark gluon matter behaves more like a deconfined quark-gluon plasma (QGP) liquid \cite{star,phenix,shuryak}. 

This fact immediately encourages some models based on either the (relativistic)
hydrodynamic \cite{bouras,romatschke}, or pure Quantum Chromodynamics (QCD)
approaches \cite{Alford}. Within the hydrodynamic framework, the plasma is
dominated by either quark \cite{romatschke} or gluon matter \cite{bouras}. In
particular, in the quark matter dominated plasma dissipative ideal hydrodynamics
has been used to fit some experimental data at high energy heavy ion program at
the Relativistic Heavy Ion Collider (RHIC) \cite{rhic}. The successful fit
requires the models to take into account very small value of the ratio shear
viscosity over entropy \cite{teaney,huovinen,kolb,kolb2,hirano,baier}. However
the puzzle must still be confirmed by the next coming experiments at the Large
Hadron Collider (LHC) \cite{lhc}. While the gluon matter plasma is motivated by
the discoveries of jet-quenching in heavy-ion-collision at RHIC indicating the
shock waves in form of March cones \cite{adams,adare}.

On the other hand, in pure QCD, QGP is described as a quark soup before
hadronization which is a phase of QCD, and exists at extremely high temperature
and/or density. It is argued that this phase consists of almost free quarks and
gluons. Therefore, the phase transition from the deconfined QGP to the hadronic
matters or vice versa gets particular interest in this approach. It
unfortunately turns out to the many body problems with large color charge which
cannot be calculated analytically using perturbation. As a result, the main
theoretical tools to explore QGP within QCD is lattice gauge theory. The
lattice calculation predicts that the phase transition occurs at approximately
175 MeV \cite{gottlieb,petreczky}. Many works on both analytical and numerical
calculations have been done at a sophisticated level providing some interesting
predictions like color glass superconductivity (CGS), color-flavor locked (CFL)
phase and so on. 

Anyway, describing the QGP as either quark or gluon matter dominated plasma
within hydrodynamic or pure QCD should be understood as conjectures in the world
of QGP. The important point is, since the QGP contains many quark-anti-quarks
and gluons, it is considerable to treat it using the well-established QCD. At
the same time from hydrodynamic point of view, the experiments strongly suggest
that QGP behaves like a fluid. In particular, in the scenario of viscous gluonic
plasma this is required to form and propagate shock waves
\cite{adler2,adams2,back}. Moreover it dissolves into an almost perfect dense
fluid \cite{zajc}. Therefore it is plausible to describe it as a strongly
interacting fluid system. In this sense, there are approaches
based on unifying or hybridizing the charge field with flow field
\cite{heinz,holm,choquet,blaizot,bistrovic,manuel}. Recently,
some works have constructed the models in a lagrangian with certain non-Abelian
gauge symmetry to the matter inside the fluid \cite{marmanis,sulaiman}. 

In this paper, we follow the scenario of gluon matter plasma as suggested by
hydrodynamic approach, while it is still governed by QCD. In this framework, the
strongly interacting system of QCD is considered as a macroscopic fluid system
rather than a result of many particles interactions. It is shown that one can
derive the total energy momentum tensor of gluon plasma within the 
model. Further, the equation of state (EoS) relevant for plasma dominated
compact star core interiors is constructed. In the model, the density ($\rho$)
and pressure ($p$) are determined and related each other through the fluid field
distribution $\phi(x)$. Hence, the EoS can be solved through the ordinary
differential equation (ODE) of $\rho$ to show the phase transition within the
model.

The paper is organized as follows. First we briefly introduce the underlying
model of gauge invariant fluid lagrangian and discuss the relevant physical
scale and region within the model. Then, the energy momentum tensor, density and
pressure in the model are derived and investigated. Subsequently it is followed
by constructing the relevant EoS for compact star core interiors. Before
summarizing the paper, the analytical solution for the pressure and density
distributions of small compact star core radius is given and discussed.

\section{The model}
\label{model}

Let us adopt the model developed by Sulaiman \etal \cite{sulaiman,tpdjun}. The
model proposes to describe the QGP as a strongly interacting gluon sea with
 quarks and anti-quarks inside. The model deploys the conventional
QCD lagrangian with SU(3) color gauge symmetry, that is,
\be 
  \L = i \bar{Q} \gamma^{\mu} \pd_\mu Q - m_Q \bar{Q} Q -
  \frac{1}{4} S^a_{\mu\nu} {S^a}^{\mu\nu} + g_s J^a_\mu {U^a}^\mu \, . 
  \label{eq:l}
\ee
Here $Q$ and $U_\mu$ represent the quark (color) triplet and gauge vector field. $g_s$ is the strong coupling constant, $J^a_\mu = \bar{Q} T^{a} \gamma_\mu Q$ and $T^a$'s belong to the SU(3) Gell-Mann matrices. The strength tensor is $S^a_{\mu\nu} = \pd_\mu U^a_\nu - \pd_\nu U^a_\mu + g_s f^{abc} U^b_\mu U^c_\nu$ with $f^{abc}$ is the structure constant of SU(3) group respectively. It should be noted that the quarks and anti-quarks feel the electromagnetic force due to the U(1) field $A_\mu$, but the size is suppressed by a factor of $e/{g_s} = \sqrt{{\alpha}/{\alpha_s}} \sim O(10^{-1})$. 

Following the original model \cite{sulaiman}, the gluon fluid is put to have a particular form in term of relativistic velocity as,
\be
	U_\mu^a = (U^a_0, \U^a) \equiv u^a_\mu \, \phi \; ,
	\label{eq:u}
\ee
with $u_\mu^a \equiv \gamma_{\v^a} (1, \v^a)$ and $\gamma_{\v^a} =  {(1 -
|\v^a|^2)}^{{-1}/2}$. $\phi = \phi(x)$ is a dimension one scalar field to keep
correct dimension and should represent the field  distribution. It is argued
that taking this form leads to the equation of motion (EOM) for a single gluon
field as follow \cite{sulaiman},
\be
	\frac{\pd}{\pd t} \left( \gamma_{\v^a} \v^a \phi \right) + \grad \left( \gamma_{\v^a} \phi \right) = -g_s \oint \d \x \, \left( \J^a_0 + F^a_0 \right) \, ,
	\label{eq:ree}
\ee
where $\J^a_\mu$ is the covariant current of gluon field, and $F^a_\mu$ is an auxiliary function which can be found in the original paper \cite{sulaiman}. It has been concluded that Eq. (\ref{eq:ree}) should be a general relativistic fluid equation, since at the non-relativistic limit Eq. (\ref{eq:ree}) coincides to the classical Euler equation. 

More precisely, this fact provides a clue that a single gluonic field $U^a_\mu$
may behave as a fluid at certain scale, beside its conventional point particle
properties with a polarization vector $\epsilon_\mu$ in the form of  $U^a_\mu =
\epsilon^a_\mu \, \phi$. One can then consider that there is a kind of ``phase
transition'',
\be
  \underbrace{\mathrm{hadronic \; state}}_{\displaystyle \epsilon^a_\mu} \longleftrightarrow \underbrace{\mathrm{QGP \; state}}_{\displaystyle u^a_\mu} \; .
\ee
As the gluon field behaves as a point particle, it is in a stable hadronic state and is characterized by its polarization vector. On the other hand in the pre-hadronic state (before hadronization) like hot QGP, the gluon field  behaves as a highly energized flow particle and the properties are dominated by its relativistic velocity. 

One should also recall that the wave function $U_\mu^a$ for a free massless particle satisfies $\pd^2 U_\mu^a = 0$ with a solution $U_\mu^a \sim \epsilon_\mu^a \, \mathrm{exp} (-i p_\nu x^\nu)$ where $p_\nu$ is the 4-momentum. Imposing the Lorentz guage condition $\pd^\mu U_\mu^a = 0$ demands $p^\mu \epsilon_\mu^a = 0$. Therefore the number of independent polarization vectors is reduced from four to three in a covariant fashion. However, one can still perform another gauge transformation to the massless $U_\mu^a$ which  makes finally only two degrees of freedom remain. Therefore, one should keep in mind that in the present model the spatial velocity has only two degrees of freedom, that means one component must be described by another two vector components. Fortunately, in real applications in cosmology or compact star, this requirement is satisfied by the assumption that the system under consideration is isotropic.

On the other hand, let us comment on the gauge transformation of velocity $u^a_\mu$. The gauge field $U_\mu^a$ is transformed as $U_\mu^a \rightarrow (U_\mu^a)^\prime = U_\mu^a + 1/{g_s} \, \pd_\mu \theta^a + f^{abc} \theta^b U_\mu^c$ with the local gauge parameter $\theta^a = \theta^a(x)$. Following the  Lorentz condition, $\theta^a$ also satisfies $\pd^2 \theta^a = 0$. If the solution is $\theta^a =  i \kappa^a \phi \, \mathrm{exp} (-i p \cdot x)$ with $\kappa^a$ constant and $\phi \sim \mathrm{exp} (-i p \cdot x)$, $u_\mu^a$ is transformed as $u_\mu^a \rightarrow (u_\mu^a)^\prime = u_\mu^a + \kappa^a/{g_s} \, p_\mu + i f^{abc} \kappa^b u_\mu^c$. The invariant 4-velocity satisfies ${u^\prime}^2 = 1 + \kappa ({m_g}/{g_s} + i f_g ) + O({\kappa}^2) \sim 1 = u^2$ since the second and third terms are suppressed by $\kappa^a \sim \theta^a \ll g_s \sim O(1)$. $f_g$ is the factor of summed colored gluon states. Therefore, one can take the form of Eq. (\ref{eq:u}) for a good approximation.

From now, throughout the paper let us focus only on the gluon sea of plasma. This means one should consider only the related gluonic terms in Eq. (\ref{eq:l}), 
\be 
  \L_g = -\frac{1}{4} S^a_{\mu\nu} {S^a}^{\mu\nu} + g_s J^a_\mu {U^a}^\mu \, . 
  \label{eq:lg}
\ee

\section{Energy momentum tensor}
\label{emt}

Now we are ready to proceed with deriving the energy momentum tensor within the model. It should be pointed out that once the hot (high energy) QGP state is achieved, the system is assumed to be predominated by the classical motion rather than the quantum effects.

Therefore the total action of matter for non-gravitational fields in a general geometry of space-time $\R$ is $S_g = \int_\R \d^4x \, \sqrt{-g} \, \L_g$, where $g$ is the determinant of metric $g_{\mu\nu}$. The variation of $S_g$ in the metric is given by $\delta S_g = -\frac{1}{2} \int_\R \d^4x \, \sqrt{-g} \, \T_{\mu\nu} \, \delta g^{\mu\nu}$. Since the energy momentum tensor density is, 
\be
  \T_{\mu\nu} = \frac{2}{\sqrt{-g}} \frac{\delta \L_g}{\delta g^{\mu\nu}} \, ,
\ee
one obtains,
\be
  \T_{\mu\nu} = S^a_{\mu\rho} {S^a}_\nu^\rho - g_{\mu\nu} \L_g + 2 g_s J^a_\mu {U^a}_\nu \, . 
  \label{eq:t}
\ee
It is clear that Eq. (\ref{eq:t}) is symmetric as expected to fulfill the Einstein gravitational EOM. The total energy momentum tensor $T_{\mu\nu}$ is given by integrating out Eq. (\ref{eq:t}) in term of total volume in the space-time under consideration. This means $T_{\mu\nu}$ is a result of bulk of gluons flow in the system.

Furthermore, in a general space-time coordinates, the components of energy momentum tensor determine the total energy density ($T_{00}$), the heat conduction ($T_{0i,i0}$), the isotropic pressure ($T_{ii}$) and the viscous stresses ($T_{ij}$ with $i \neq j$) of the gluonic plasma. Of course, in this case the derivative $\pd_\mu$ inside the strength tensor $S_{\mu\nu}$ should be replaced by the covariant one, $\grad_\mu$. Also, the energy momentum tensor satisfies the conservation condition, $\grad_\mu \, \T^{\mu\nu} = 0$. Nevertheless, one can trivially conclude that the model induces non-zero viscosity since generally $T_{ij} \neq 0$ for $i \neq j$. From the experimental clues, however the size should be small such that it is always treated perturbatively in most hydrodynamics models \cite{teaney,huovinen,kolb,kolb2,hirano,baier}.

Before going further to apply these results, one should determine the quark
current $J^a_\mu$ in Eq. (\ref{eq:t}). This can be simply calculated by
considering the EOM (Dirac equation) of a single colored quark ($Q$) or
anti-quark ($\bar{Q}$) with 4-momentum $p_\mu$. Since the solution of the EOM is
$Q(p,x) = q(p) \, \exp(-i p \cdot x)$, one immediately gets $\bar{q} \gamma_\mu
q = 4 p_\mu$. Assuming that all colored quarks / anti-quarks have the same
momenta and  the velocity of gluons and quarks inside the gluon sea are
homogeneity, approximately $J^a_\mu {U^a}^\mu \propto 4 p_\mu U^\mu = 4 m_Q
\phi$ since $u_\mu u^\mu = u^2 = 1$.

\section{Equation of state}
\label{eos}

Now we are ready to consider the compact star interiors in the model, particularly before transforming itself into neutron star.

The whole compact star is commonly described as a static spherically symmetric
space-time represented by Schwarzschild geometry. This means one deals with the
relativistic gravitational equations for the interior of spherically symmetric
plasma distribution in the core. In the region under consideration the presence
of gluonic fields flow induces non-zero energy momentum tensor which is making
up the star. This is the phase before the neutron star is getting mature.
Starting from the stellar nebula made of hot plasma which is gradually getting
colder as the hadronization occurs from the colder surface, while the inner core
is still in pure hot plasma state.

As a consequence of the diagonal metric of Schwarzschild space-time, the model falls back to the perfect fluid without viscosity and heat conduction, \ie $T_{0i} = T_{ij} = 0$ for $i \neq j$. Also, since the plasma distribution should be spherically isotropic, it is considerable to put $v_1 = v_2 = v_3 = v$ as constant for all colored gluons. This assumption is consistent with the degree of freedom counting discussed in the preceding section. Moreover, the vanishing off-diagonal components of the Ricci tensor, $R_{i0}$, actually forces the spatial 3-velocity of the fluid must vanish everywhere. Hence particular assumption for $v_i$ is indeed not necessary. However, the gluon distribution still depends on the radius length, $\phi = \phi(r)$. 

For the sake of simplicity one can put homogeneous gluon fields for all color states, \ie $U_\mu^a = U_\mu$ for all $a = 1, \cdots, 8$. This yields,
\bea
  \T_{\mu\nu} & = & \left[ 8 \, g_s \, f_Q \, m_Q \, \phi(r) + g_s^2 \, f_g^2 \, \phi(r)^4 \right] u_\mu u_\nu  
  \nonumber\\
  && - \left[ 4 \, g_s \, f_Q \, m_Q \, \phi(r) - \frac{1}{4} g_s^2 \, f_g^2 \, \phi(r)^4 \right] g_{\mu\nu} \, ,
  \label{eq:tf}
\eea
where $f_g$ is the factor of summed colored gluon states from the structure constant $f^{abc}$, while $f_Q$ is the factor of summed colored quark states from $J^a_\mu {U^a}^\mu$. Remind that the energy momentum tensor for perfect fluid takes the form,
\be
  \T_{\mu\nu} = \left( \E + \P \right) u_\mu u_\nu  - \P \, g_{\mu\nu} \, .
  \label{eq:tpf}
\ee
Here $\E$ and $\P$ denote the density and isotropic pressure for single fluid field, each is related to the total density and pressure of the system through $\rho = \oint \d^4 x \, \E$ and $P = \oint \d^4 x \, \P$ respectively.
Obviously, from Eqs. (\ref{eq:tf}) and (\ref{eq:tpf}) one can obtain the density and pressure in the model as follows,
\bea
  P(r) & = & \int^{\beta_s}_{\beta_0} \d t \int  \d V \left[ 4 \, g_s \, f_Q \, m_Q \, \phi(r) - \frac{1}{4} g_s^2 \, f_g^2 \, \phi(r)^4 \right] 
  \nonumber\\
  & = & \frac{4 \, g_s \, f_Q \, m_Q}{T} \int  \d V \left[ 1 - \frac{g_s \, f_g^2}{16 \, f_Q \, m_Q}  \phi(r)^3 \right] \phi(r) \, , 
  \label{eq:p}\\
  \rho(r) & = & \int^{\beta_s}_{\beta_0} \d t \int  \d V \left[ 4 \, g_s \, f_Q \, m_Q \, \phi(r) + \frac{5}{4} g_s^2 \, f_g^2 \, \phi(r)^4 \right] 
  \nonumber\\
  & = & \frac{4 \, g_s \, f_Q \, m_Q}{T} \int  \d V \left[ 1 + \frac{5 \, g_s \, f_g^2}{16 \, f_Q \, m_Q}  \phi(r)^3 \right] \phi(r) \, , 
  \label{eq:e}
\eea
at a finite temperature $\beta = 1/{T}$ in a 3-dimensional spatial volume $V$.
Here, $T_s$ and $T_0$ denote the core surface and inner core temperatures.

The proper spatial volume element for Schwarzschild geometry is $\d V = \sqrt{B(r)} r^2 \, \sin\theta \, \d r \, \d \theta \, \d \varphi$ with radius $r$ and two angles $\theta$ and $\varphi$ in spherical coordinates. The solution for $B(r)$ is given by $B(r) = \left[ 1 - {2 G m(r)}/{r} \right]^{-1}$ with $m(r) = 4 \pi \int^r_0 \d\bar{r} \rho(\bar{r}) \, \bar{r}^2$ is the 'bare mass'. This generates the proper integrated mass $\tilde{m}(r)$ contained within a coordinate radius $r$ inside the star. Anyway, the inner structure of star with Schwarzschild geometry is well known as Tolman-Oppenheimer-Volkoff (TOV) equation which relates  density and  pressure in a unique way \cite{tolman,ov}.

Obviously, using Eqs. (\ref{eq:e}) and (\ref{eq:p}) one can construct certain
EoS,
\be
  P(r) = w(r) \, \rho(r) \; ,
\ee
where in contrast with the conventional cosmological models, 
\be
  w(r) \equiv \frac{\displaystyle \int  \d V \left[ 1 - \frac{g_s \, f_g^2}{16 \, f_Q \, m_Q}  \phi(r)^3 \right] \phi(r)}{\displaystyle \int  \d V \left[ 1 + \frac{5 \, g_s \, f_g^2}{16 \, f_Q \, m_Q}  \phi(r)^3 \right] \phi(r)} \; ,
\ee
is not a constant. This is actually one of the important consequences in the
present model. Once the field distribution $\phi(r)$ is determined one can
obtain certain forms of density and pressure.

\section{Pressure and density distribution}
\label{pdb}

For the sake of convenience later on, let us redefine the radius $r$ to be the
dimensionless one, \ie the ratio of core and compact star radius : $r
\rightarrow \rp \equiv r/\rs$. Here $\rs$ is the compact star radius. Then, the
pressure and density in Eqs. (\ref{eq:e}) and (\ref{eq:p}) can be expressed as,
\bea
  F_X(\rp) & = & {F_X}_0 + 16 \pi \, g_s \, f_Q \, m_Q \, \rs^3 \frac{T_0 -
T_s}{T_0 T_s} \nonumber \\
  && \times 
  \int_0^\rp \d \rp \rp^2 \sqrt{B(\rp)} \left[ 1 + k_X \frac{g_s f_g^2}{16 \, f_Q \, m_Q} \phi(\rp)^3 \right] \phi(\rp) \; ,
\eea
after integrating out the time component. $F_X$ denotes the pressure $P$ or density $\rho$ for $X = P, \rho$ respectively, while $k_P = -1$ and $k_\rho = 5$. ${F_X}_0$ represents the initial $F_X$. Taking $A_1(\rp) = \int_0^\rp \d \rp \rp^2 \sqrt{B(\rp)} \phi(\rp)$, $A_2 = \int_0^\rp \d \rp \rp^2 \sqrt{B(\rp)} \phi(\rp)^4$, $k_1 = 16 \pi \, g_s \, f_Q \, m_Q \, \rs^3$ and $k_2 = \pi \, g_s^2 \, f_g \, \rs^3$ yield, 
\be
  F_X(\rp) = {F_X}_0 + \frac{T_0 - T_s}{T_0 T_s} \left[ k_1 \, A_1(\rp) + k_X k_2 \, A_2(\rp) \right] \; .
  \label{eq:pr} 
\ee

Finally, by substituting the expressions for $m(\rp)$ and $B(\rp)$ into Eq. (\ref{eq:pr}) one finds,
\be
  F_X(\rp) = {F_X}_0 +  \frac{T_0 - T_s}{T_0 T_s} 
      \int_0^\rp {\d \rp \frac{\left[ k_1 + k_X k_2 \phi(r)^3 \right] \phi(\rp)}{\displaystyle \sqrt{1 - \frac{8 \pi G}{\rp} \int_0^{\rp} \d \bar{\rp} \rho(\bar{\rp})  \bar{\rp}^2}}} \; .
  \label{eq:fx}
\ee
Taking its derivative in term of $\rp$, Eq. (\ref{eq:fx}) leads to an ODE as below,
\be
  \Lambda_1(\rp) F_X^{\prime\prime}(\rp) + T^2 \, {F_X^\prime(\rp)}^3 - \Lambda_2 T^2 \rp^2 \, F_X(\rp) {F_X^\prime(\rp)}^3 - 
  \Lambda_3(\rp) F_X^\prime(\rp) = 0 \; ,
  \label{eq:ode}
\ee
where $T \equiv {T_0 T_s}/{(T_0 - T_s)}$ and,
\bea
  \Lambda_1(\rp) & = & 2 \rp^5 \phi(\rp)^2 \left[ k_1 + k_X k_2 \phi(\rp)^3 \right]^2 \; , \\
  \Lambda_2 & = & 8 \pi G \; , \\
  \Lambda_3(\rp) & = & \left[ 5 \rp^4 \phi(\rp)^2 + 2 \rp^5 \phi(\rp) \phi^\prime(\rp) \right] 
      \left[ k_1 + k_X k_2 \phi(\rp)^3 \right]^2 
    \nonumber \\
    && + 6 k_X k_2 \, \rp^5 \left[ k_1 + k_X k_2 \phi(\rp)^3 \right] \phi(\rp)^4 \phi^\prime(\rp)  \; .
\eea
Note that a prime in $\phi$ or $F_X$ means a derivative in term of $\rp$.

Nevertheless, one can also derive a ``differential'' EoS by taking the
derivative of Eq. (\ref{eq:fx}) in term of $\rp$ and eliminating $B(\rp)$, that
is,
\be
  P^\prime(\rp) = \frac{16 \, f_Q \, m_Q - g_s f_g^2 \, \phi(\rp)^3}{16 \, f_Q \, m_Q + 5 g_s f_g^2 \, \phi(\rp)^3} \, \rho^\prime(\rp) \; .
\ee

Now let us solve the ODE in Eq. (\ref{eq:ode}). The ODE can be solved
analytically by taking an approximation of small compact star core, \ie $\rp \ll
1$. This approximation is quite natural since by definition the compact star
cores should be small enough. Hence one can expand the ODE near the origin ($\rp
\rightarrow 0$) to obtain the solution order by order, 
\be
  F_X(\rp) \sim \left. F_X(\rp) + F_X^{(\MakeUppercase{\romannumeral 1})}(\rp)
\rp + \frac{1}{2!} F_X^{(\MakeUppercase{\romannumeral 2})}(\rp) \rp^2 +
\frac{1}{3!} F_X^{(\MakeUppercase{\romannumeral 3})}(\rp) \rp^3 + \cdots
\right|_{\rp=0} \; .
\ee
First of all, taking the first derivative on  Eq. (\ref{eq:ode}) in term of
$\rp$ yields,
\be
  3 T^2 {F_X^{(\MakeUppercase{\romannumeral 1})}}^2(0)
F_X^{(\MakeUppercase{\romannumeral 2})}(0) = 0 \; ,
  \label{eq:1st}
\ee
since $\Lambda_1^\prime(0) = \Lambda_3^\prime(0) = 0$. Further derivatives up to 7th order give,
\bea
  F_X^{(\MakeUppercase{\romannumeral 3})}(0) & = & \frac{1}{3
F_X^{(\MakeUppercase{\romannumeral 1})}(0)^2} \left[ 2 \Lambda_2 \, F_X(0)
F_X^{(\MakeUppercase{\romannumeral 1})}(0)^3 - 6
F_X^{(\MakeUppercase{\romannumeral 1})}(0) F_X^{(\MakeUppercase{\romannumeral
2})}(0)^2 \right] \; , 
  \label{eq:2nd}\\
  F_X^{(\MakeUppercase{\romannumeral 4})}(0) & = &
\frac{1}{F_X^{(\MakeUppercase{\romannumeral 1})}(0)^2} \left[ 6 \Lambda_2 F_X(0)
F_X^{(\MakeUppercase{\romannumeral 1})}(0)^2 F_X^{(\MakeUppercase{\romannumeral
2})}(0) + 2 \Lambda_2 F_X^{(\MakeUppercase{\romannumeral 1})}(0)^4 \right.
  \nonumber \\
  && \left. - 2 T^2 F_X^{(\MakeUppercase{\romannumeral 2})}(0)^3 - 6
F_X^{(\MakeUppercase{\romannumeral 1})}(0) F_X^{(\MakeUppercase{\romannumeral
2})}(0) F_X^{(\MakeUppercase{\romannumeral 3})}(0) \right] \; , \\
  F_X^{(\MakeUppercase{\romannumeral 5})}(0) & = & \frac{1}{3 T^2
F_X^{(\MakeUppercase{\romannumeral 1})}(0)^2} \left[
\Lambda_3^{(\MakeUppercase{\romannumeral 4})}(0)
F_X^{(\MakeUppercase{\romannumeral 1})}(0) + 36 \Lambda_2 T^2 F_X(0)
F_X^{(\MakeUppercase{\romannumeral 1})}(0)^2 F_X^{(\MakeUppercase{\romannumeral
3})}(0) \right.
  \nonumber\\
  && \left. + 72 \Lambda_2 T^2 F_X(0) F_X^{(\MakeUppercase{\romannumeral 1})}(0)
F_X^{(\MakeUppercase{\romannumeral 2})}(0)^2 + 82 \Lambda_2 T^2
F_X^{(\MakeUppercase{\romannumeral 1})}(0)^3 F_X^{(\MakeUppercase{\romannumeral
2})}(0)\right.
  \nonumber\\
  && \left. - 24 T^2 F_X^{(\MakeUppercase{\romannumeral 1})}(0)
F_X^{(\MakeUppercase{\romannumeral 2})}(0) F_X^{(\MakeUppercase{\romannumeral
4})}(0) 
  - 18 T^2 F_X^{(\MakeUppercase{\romannumeral 1})}(0)
F_X^{(\MakeUppercase{\romannumeral 3})}(0)^2\right.
  \nonumber\\
  && \left.  - 36 T^2
F_X^{(\MakeUppercase{\romannumeral 2})}(0)^2 F_X^{(\MakeUppercase{\romannumeral
3})}(0) \right] \; , \\
  F_X^{(\MakeUppercase{\romannumeral 6})}(0) & = & \frac{1}{3 T^2
F_X^{(\MakeUppercase{\romannumeral 1})}(0)^2} \left[ 
  76 \Lambda_2 T^2 F_X^{(\MakeUppercase{\romannumeral 1})}(0)^3
F_X^{(\MakeUppercase{\romannumeral 3})}(0) + 36 \Lambda_2 T^2 F_X(0)
F_X^{(\MakeUppercase{\romannumeral 1})}(0)^2 \right.
  \nonumber\\
  && \left.  + 24 \Lambda_2 T^2 F_X(0) F_X^{(\MakeUppercase{\romannumeral
1})}(0)^2  F_X^{(\MakeUppercase{\romannumeral 4})}(0) + 522 \Lambda_2 T^2
F_X^{(\MakeUppercase{\romannumeral 1})}(0)^2 F_X^{(\MakeUppercase{\romannumeral
2})}(0)^2 \right.
  \nonumber\\
  && \left. + 84 \Lambda_2 T^2  F_X(0) F_X^{(\MakeUppercase{\romannumeral
2})}(0)^3 + 188 \Lambda_2 T^2  F_X^{(\MakeUppercase{\romannumeral 1})}(0)^3
F_X^{(\MakeUppercase{\romannumeral 3})}(0) \right.
  \nonumber\\
  && \left. + 342 \Lambda_2 T^2  F_X(0) F_X^{(\MakeUppercase{\romannumeral
1})}(0) F_X^{(\MakeUppercase{\romannumeral 2})}(0)
F_X^{(\MakeUppercase{\romannumeral 3})}(0) \right.
  \nonumber\\
  && \left.+ 36 \Lambda_2 T^2 F_X(0)
F_X^{(\MakeUppercase{\romannumeral 1})}(0)^2 F_X^{(\MakeUppercase{\romannumeral
4})}(0)  + \Lambda_3^{(\MakeUppercase{\romannumeral 5})}(0)
F_X^{(\MakeUppercase{\romannumeral 1})}(0)\right.
  \nonumber\\
  && \left. + 5
\Lambda_3^{(\MakeUppercase{\romannumeral 4})}(0)
F_X^{(\MakeUppercase{\romannumeral 2})}(0) -
\Lambda_1^{(\MakeUppercase{\romannumeral 5})} F_X^{(\MakeUppercase{\romannumeral
2})}(0) 
  - 72 T^2 F_X^{(\MakeUppercase{\romannumeral 2})}(0)
F_X^{(\MakeUppercase{\romannumeral 3})}(0)^3 \right.
  \nonumber\\
  && \left. - 60 T^2 F_X^{(\MakeUppercase{\romannumeral 2})}(0)^2
F_X^{(\MakeUppercase{\romannumeral 4})}(0) - 18 T^2
F_X^{(\MakeUppercase{\romannumeral 2})}(0) F_X^{(\MakeUppercase{\romannumeral
3})}(0)^2 \right.
  \nonumber\\
  && \left. - 60 T^2 F_X^{(\MakeUppercase{\romannumeral 1})}(0)
F_X^{(\MakeUppercase{\romannumeral 3})}(0) F_X^{(\MakeUppercase{\romannumeral
4})}(0) \right.
  \nonumber\\
  && \left.  - 30 T^2 F_X^{(\MakeUppercase{\romannumeral 1})}(0)
F_X^{(\MakeUppercase{\romannumeral 2})}(0) F_X^{(\MakeUppercase{\romannumeral
5})}(0)
  \right] \; ,\\
  F_X^{(\MakeUppercase{\romannumeral 7})}(0) & = & \frac{1}{3 T^2
F_X^{(\MakeUppercase{\romannumeral 1})}(0)^2} \left[ 
  378 \Lambda_2 T^2 F_X^{(\MakeUppercase{\romannumeral 1})}(0)^3
F_X^{(\MakeUppercase{\romannumeral 4})}(0) \right.
  \nonumber\\
  && \left.  +
66 \Lambda_2 T^2 F_X(0) F_X^{(\MakeUppercase{\romannumeral 1})}(0)^2
F_X^{(\MakeUppercase{\romannumeral 5})}(0) +
\Lambda_3^{(\MakeUppercase{\romannumeral 6})}(0)
F_X^{(\MakeUppercase{\romannumeral 1})}(0) \right.
  \nonumber\\
  && \left. + 15
\Lambda_3^{(\MakeUppercase{\romannumeral 4})}(0)
F_X^{(\MakeUppercase{\romannumeral 3})}(0) - 6
\Lambda_1^{(\MakeUppercase{\romannumeral 5})}(0)
F_X^{(\MakeUppercase{\romannumeral 3})}(0) \right.
  \nonumber\\
  && \left.  - 72 T^2
F_X^{(\MakeUppercase{\romannumeral 3})}(0)^4 - 18 T^2
F_X^{(\MakeUppercase{\romannumeral 3})}(0)^3 - 60 T^2
F_X^{(\MakeUppercase{\romannumeral 1})}(0) F_X^{(\MakeUppercase{\romannumeral
4})}(0)^2  \right.
  \nonumber\\
  && \left. - 90 T^2 F_X^{(\MakeUppercase{\romannumeral 1})}(0)
F_X^{(\MakeUppercase{\romannumeral 3})}(0)
F_X^{(\MakeUppercase{\romannumeral 5})}(0) 
 \right] \; .
  \label{eq:7th}
\eea

\begin{figure}[t]
        \centering 
	\includegraphics[width=11cm,angle=0,trim=0 0 0 0]{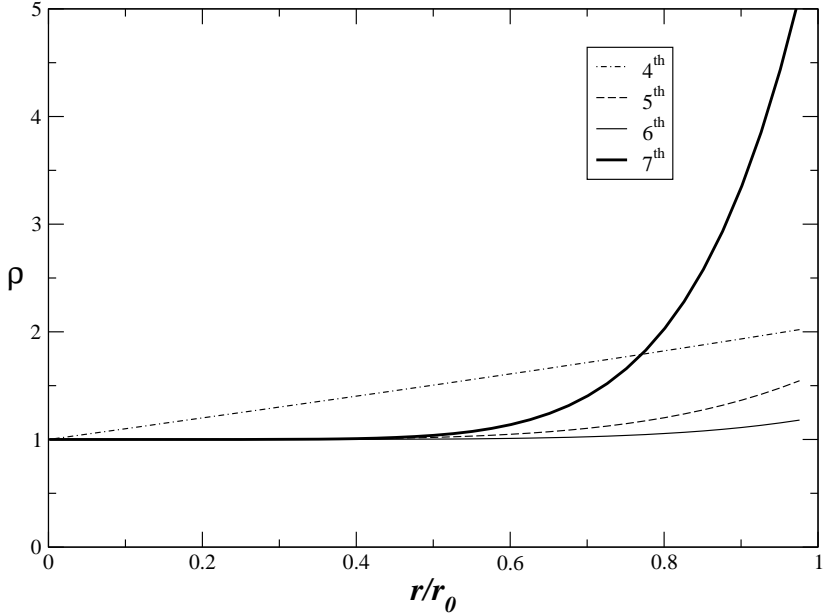}
        \caption{The density distribution as a function of the normalized
compact star core radius with $T_s = 175$ MeV and $T_0 = 1$ GeV up to
4$^\mathrm{th}$ (dotted-dashed line), 5$^\mathrm{th}$ (dashed line),
6$^\mathrm{th}$ (thin solid line) and 7$^\mathrm{th}$ (thick solid line) order
expansions.}
        \label{fig:d-r}
\end{figure}

\begin{figure}[t]
        \centering 
	\includegraphics[width=11cm,angle=0,trim=0 0 0 0]{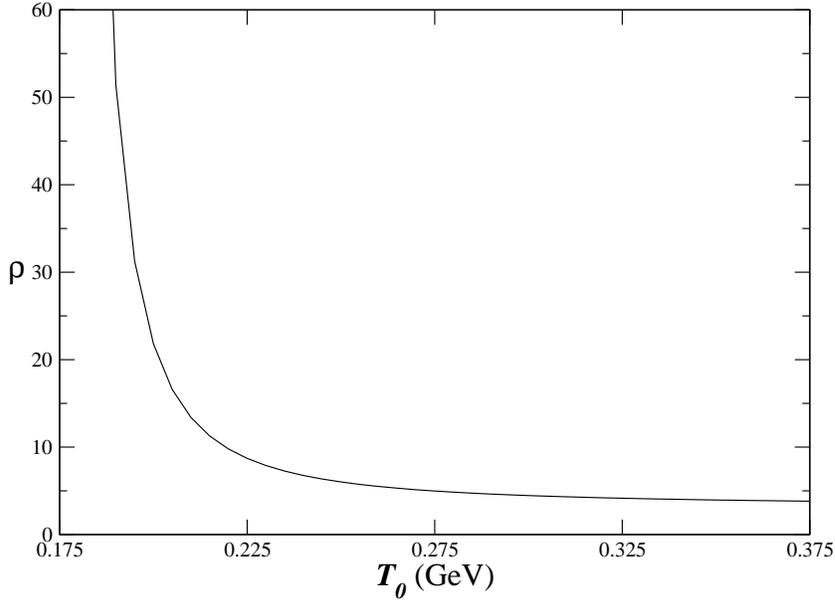}
        \caption{The density distribution as a function of inner temperature
with the normalized star radius $\rp = 0.01$ and $T_s = 175$ MeV.}
        \label{fig:d-t}
\end{figure}

From Eq. (\ref{eq:1st}), non-trivial solution is obtained for
$F_X^{(\MakeUppercase{\romannumeral 2})}(0) = 0$ which leads to
$F_X^{(\MakeUppercase{\romannumeral 1})}(0) = \mathrm{const}$. Without loss of
generality, one can choose $F_X(0) = 1$. These results are substituted
subsequently into Eqs. (\ref{eq:2nd})$\sim$(\ref{eq:7th}) to get the complete
solution. The solution is depicted in Figs. \ref{fig:d-r} and \ref{fig:d-t} for
the density distribution, \ie $F_X = \rho$. 

It is important to derive the solution analytically to investigate the behavior
of phase transition in term of higher order solutions. The results of total
density as a function of $r^\prime$ are depicted in Fig. \ref{fig:d-r} for the
solution up to 4$^\mathrm{th}$ (dotted-dashed line), 5$^\mathrm{th}$ (dashed
line),
6$^\mathrm{th}$ (thin solid line) and 7$^\mathrm{th}$ (thick solid line) order
expansions. It is clear that the phase transtion occurs at 7$^\mathrm{th}$ order
accuracy. In particular the contour is mainly governed by the dissipative term, 
$\Lambda_3^{(\MakeUppercase{\romannumeral 6})}(0)$. Those results are obtained
by taking a particular form of distribution function, $\phi(r) \sim \exp (a
r)$ with $a = 0.4$.

\section{Summary}

The QGP, in particular gluon matter, dominated compact start core interior has
been investigated using the fluid QCD lagrangian. From the lagrangian one can
derive the energy-momentum tensor and subsequently the density and pressure
distributions from the first principle. In the model, both density and pressure
are related each other and form an extended EoS of perfect fluid through the
field distribution $\phi(r)$. The relation leads to the linear EoS known in the
conventional cosmological model, but allowing the EoS parameter $w$ to be a
function of temperature $T$ and core radius $r$. The analytical solutions for
the total pressure and density distribution equations are also given for small
normalized compact star core radius, $\rp \ll 1$. 

The model describes the dense gluon plasma inside the compact star core as a
bulk dynamics of gluonic matters before the cooling phase near the compact star
core surface as depicted in Fig. \ref{fig:d-r}. It should be noted that the
figure provides a general contour for pressure or density distribution inside
the core. This suggests that near its surface the hadronization is getting
involved to form surfaces of small compact star cores,

On the other hand, the phase transition for certain size of compact star core in
term of temperature changing is shown in Fig. \ref{fig:d-t}. It can be observed 
that the phase transition occurs at the scale of hadronization around the core
surface, while it keeps fluidity as the temperature is getting higher in the
core interiors.

\begin{acknowledgements}
The authors thank M. K. Nurdin for fruitful discussion during the final stage of
this paper. This work is funded by Riset Kompetitif LIPI in fiscal year 2011
under Contract no.  11.04/SK/KPPI/II/2011. 
\end{acknowledgements}

\bibliographystyle{spphys}
\bibliography{compactstar}

\begin{thebibliography}{10}
\providecommand{\url}[1]{{#1}}
\providecommand{\urlprefix}{URL }
\expandafter\ifx\csname urlstyle\endcsname\relax
  \providecommand{\doi}[1]{DOI \discretionary{}{}{}#1}\else
  \providecommand{\doi}{DOI \discretionary{}{}{}\begingroup
  \urlstyle{rm}\Url}\fi

\bibitem{star}
{C. Adler \textit{et.al.} (STAR Collaboration)}, Phys. Rev. C \textbf{66},
  034904 (2002)

\bibitem{phenix}
{K. Adeox \textit{et.al.} (PHENIX Collaboration)}, Nucl. Phys. A \textbf{757},
  184 (2005)

\bibitem{shuryak}
E.~Shuryak, Nucl.Phys. A \textbf{774}, 387 (2006)

\bibitem{bouras}
I.~Bouras, E.~Molnar, H.~Niemi, Z.~Xu, A.~El, O.~Fochler, C.~Greiner,
  D.~Rischke, Phys. Rev. Lett. \textbf{103}, 032301 (2009).
\newblock \doi{10.1103/PhysRevLett.103.032301}

\bibitem{romatschke}
P.~Romatschke, Int. J. Mod. Phys. E \textbf{E19}, 1 (2010)

\bibitem{Alford}
M.G. Alford, K.~Rajagopal, T.~Schaefer, A.~Schmitt, Rev. Mod. Phys.
  \textbf{80}, 1455 (2008).
\newblock \doi{10.1103/RevModPhys.80.1455}

\bibitem{rhic}
M.~Harrison, T.~Ludlam, S.~Ozaki, Nucl. Inst. Meth. Phys. Res. A \textbf{499},
  235 (2003)

\bibitem{teaney}
D.~Teaney, J.~Lauret, E.V. Shuryak, Phys. Rev. Lett. \textbf{86}, 4783 (2001)

\bibitem{huovinen}
P.~Huovinen, P.F. Kolb, U.W. Heinz, P.V. Ruuskanen, S.A. Voloshin, Phys. Lett.
  B \textbf{503}, 58 (2001)

\bibitem{kolb}
P.F. Kolb, U.W. Heinz, P.~Huovinen, K.J. Eskola, K.~Tuominen, Nucl. Phys. A
  \textbf{696}, 197 (2001)

\bibitem{kolb2}
P.F. Kolb, R.~Rapp, Phys. Rev. C \textbf{67}, 044903 (2003)

\bibitem{hirano}
T.~Hirano, K.~Tsuda, Phys. Rev. C \textbf{66}, 054905 (2002)

\bibitem{baier}
R.~Baier, P.~Romatschke, Eur. Phys. J. C \textbf{51}, 677 (2007)

\bibitem{lhc}
J.~Jowett, {LHC Lead Ion Beam Commissioning in LHC Design Report}.
\newblock Tech. rep., CERN (2009)

\bibitem{adams}
{J. Adams, et al. (STAR Collaboration)}, Phys. Rev. Lett. \textbf{91}, 172302
  (2003)

\bibitem{adare}
{A. Adare, et al. (PHENIX Collaboration)}, Phys. Rev. Lett. \textbf{101},
  232301 (2008).
\newblock \doi{10.1103/PhysRevLett.101.232301}

\bibitem{gottlieb}
S.~Gottlieb, J. Phys. Conf. Ser. \textbf{78}, 012023 (2007)

\bibitem{petreczky}
P.~Petreczky, Europ. Phys. J. Special Topics \textbf{155}, 1951 (2008)

\bibitem{adler2}
{S.S. Adler, et al. (PHENIX Collaboration)}, Phys.Rev.Lett \textbf{91}, 182301
  (2003)

\bibitem{adams2}
{J. Adams, et al. (STAR Collaboration)}, Phys.Rev.Lett. \textbf{92}, 052302
  (2004)

\bibitem{back}
{B.B. Back et.al. (Phobos Collaboration)}, Phys.Rev.C \textbf{72}, 051901
  (2005)

\bibitem{zajc}
W.A. Zajc, Nucl. Phys. A \textbf{805}, 283 (2008)

\bibitem{heinz}
U.~Heinz, Phys. Rev. Lett. \textbf{51}, 351 (1983)

\bibitem{holm}
D.D. Holm, B.A. Kupershmidt, Phys. Rev. D \textbf{30}, 2557 (1984)

\bibitem{choquet}
Y.~Choquet-Bruhat, J. Math. Phys. \textbf{33}, 1782 (1992)

\bibitem{blaizot}
J.P. Blaizot, E.~Iancu, Nucl. Phys. B \textbf{421}, 565 (1994)

\bibitem{bistrovic}
B.~Bistrovic, R.~Jackiw, H.~Li, V.P. Nair, S.Y. Pi, Phys. Rev. D \textbf{67},
  025013 (2003)

\bibitem{manuel}
C.~Manuel, S.~Mrowczynski, Phys. Rev. D \textbf{74}, 105003 (2006)

\bibitem{marmanis}
Marmanis, Phys. of Fluid \textbf{10}, 1428 (1998)

\bibitem{sulaiman}
A.~Sulaiman, A.~Fajarudin, T.P. Djun, L.T. Handoko, Int. J. Mod. Phys. A
  \textbf{24}, 3630 (2009).
\newblock \doi{10.1142/S0217751X09047284}

\bibitem{tpdjun}
T.P. Djun, L.T. Handoko, in \emph{Proceeding of the Conference in Honour of
  Murray Gell-Mann's 80th Birthday : Quantum Mechanics, Elementary Particles,
  Quantum Cosmology and Complexity} (2011), pp. 419--425.
\newblock \doi{10.1142/9789814335614\_0040}

\bibitem{tolman}
R.C. Tolman, Proc. Nat. Acad. Sci. \textbf{20}, 169 (1934)

\bibitem{ov}
J.R. Oppenheimer, G.M. Volkoff, Phys. Rev. \textbf{55}, 374 (1939)

\end{thebibliography}

\end{document}